\documentclass[10pt]{iopart}    
\usepackage{graphicx}
\usepackage{color}

\newcommand{\dcr}{DC-readout}

\begin{document}
\title{DC-readout of a signal-recycled gravitational wave detector}
\author{S.~Hild$^1$, H.~Grote$^2$, J.~Degallaix $^2$,  S.~Chelkowski$^1$, K.~Danzmann$^2$,
 A.~Freise$^1$,
 M.~Hewitson$^2$, J.~Hough$^3$, H.~L\"uck$^2$, M.~Prijatelj$^2$, K.A.~Strain$^3$, J.R.~Smith$^4$, B.~Willke$^2$}
\address{$^1$ School of Physics and Astronomy, University of
Birmingham,   Edgbaston, Birmingham, B15 2TT, UK}
\address{$^2$ Max-Planck-Institut f\"ur Gravitationsphysik
(Albert-Einstein-Institut) and Leibniz Universit\"at Hannover,
Callinstr.~38, D--30167 Hannover, Germany.}
\address{$^3$ SUPA, Physics \& Astronomy, University of Glasgow,
 Glasgow G12 8QQ, Great Britain}
\address{$^4$ Syracuse University, Department of Physics, 201 Physics Building, Syracuse, New
York 13244-1130, USA.}
 \ead{hild@star.sr.bham.ac.uk}

\begin{abstract}
All first-generation large-scale gravitational wave
detectors are operated at the dark fringe and use a heterodyne
readout employing radio frequency (RF) modulation-demodulation techniques. However,
the experience in the currently running interferometers reveals
several problems connected with a heterodyne readout, of which phase
noise of the RF modulation is the most serious one. A homodyne
detection scheme (DC-readout), using the highly stabilized and
filtered carrier light as local oscillator for the readout, is
considered to be a favourable alternative. Recently a DC-readout
scheme was implemented on the GEO\,600 detector. We describe the
results of first measurements and give a comparison of the
performance achieved with homodyne and heterodyne readout. 
The implications of the combined use of DC-readout and signal recycling
are considered.

\end{abstract}


\section{Introduction}

So far all of the large-scale gravitational wave
detectors LIGO \cite{ligo}, Virgo \cite{virgo}, TAMA300 \cite{tama} and
GEO\,600 \cite{geo} have been operated at the dark fringe using a heterodyne
 readout technique.
 However, the operation of the currently running interferometers
revealed several problems  being connected with heterodyne readout.
Changing the readout system to \emph{DC-readout}, which is
a special case of homodyne detection, can
 be beneficial for future gravitational wave detectors. Therefore it is planned to 
implement DC-readout in Enhanced LIGO \cite{eligo} and Virgo+ \cite{virgo+}
 as well as in Advanced LIGO \cite{aligo, aligo2} and Advanced Virgo \cite{adv}.

In this article we give an overview of work related to DC-readout carried
out at the GEO\,600 gravitational wave detector.
We begin, in Section~\ref{sec:principle}, with a brief description of the
principles of heterodyne, homodyne and DC-readout applied to a simple Michelson interferometer
as an illustrative example.
 In Section~\ref{sec:advantages} we give a brief summary of the
 expected advantages and disadvantages of DC-readout over heterodyne detection.
  We compare the simulated shot-noise limited sensitivity of GEO\,600 with heterodyne
 and homodyne readout in Section~\ref{sec:shotnoise}.
 It is found that in the case of detuned signal recycling not only is the
overall level of the shot noise different for homodyne and heterodyne readout, but also
 the shape of the optical response.
The
 actual experimental scheme for realisation of DC-readout in the GEO\,600 interferometer
is described in detail in Section~\ref{sec:realisation}.
 As we show in Section~\ref{sec:rotation} the simulated
 shape change of the optical response is accurately confirmed by experimental observations.
A comparison of the actual sensitivity of the GEO\,600 detector for homodyne and
 the nominal heterodyne detection scheme is given in Section~\ref{sec:sensitivity}.
Finally, Section~\ref{sec:summary} provides a summary of this article and an outlook.

\section{Definitions: Heterodyne, Homodyne and DC-readout}
\label{sec:principle}

\begin{figure}[Htb]
\centering
\includegraphics[width=0.85\textwidth]{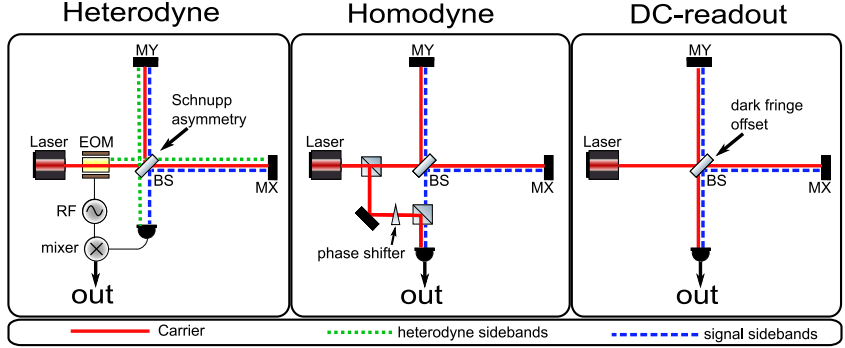}
\caption{Illustration of three different readout methods of a Michelson interferometer:
heterodyne, homodyne and DC-readout. (A detailed explanation is given in the text.)  }
\label{fig:principle}
\end{figure}

Figure~\ref{fig:principle} shows simplified schematics of three different readout
methods applied to a basic Michelson interferometer. Usually Michelson
interferometers used for gravitational wave detection are operated at a dark
 fringe\footnote{Operating at the dark fringe has the advantage
 of providing good suppression of common mode noise and allows to 
make use of power recycling.}:
The differential arm-length is controlled
to give destructive interference at the output port, i.e.\ ideally
 no carrier light ($f_{\rm c}$, red solid line) reaches the photo detector.
 The interaction of a gravitational wave with the Michelson interferometer can 
be considered as shorting of one interferometer arm, whilst the perpendicular one 
is elongated. This change of the differential arm length causes phase modulation
sidebands, i.e. gravitational
wave signal sidebands (blue dashed line). In contrast to the carrier light the gravitational
wave signal sidebands
 interfere constructively at the beam splitter, exit the interferometer at its output port
and finally reach the photo detector.
The absolute frequency of the gravitational signal sidebands given
 is by $f_{\rm sig} =  f_{\rm c} \pm f_{\rm gw}$, where $f_{\rm gw}$ is the
 frequency of the gravitational wave (usually in the audio-band) and $f_{\rm c}$ 
the frequency of the main laser light (carrier).
Since $f_{\rm sig}$ is
a few hundred Terahertz, the photo diode cannot directly detect
the gravitational wave signal, unless the presence of an optical local oscillator
is provided. Heterodyne, homodyne and DC-readout use different concepts to
ensure the presence of a low-noise optical local oscillator at the output port photo diode.

In the heterodyne schemes, commonly used by the first generation
gravitational wave detectors,
 radio frequency (RF) sidebands ($f_{\rm het}$, green dotted lines)
are modulated onto the light at the input of
the Michelson interferometer (Schnupp modulation \cite{schnupp}).
Introducing a macroscopic arm length difference of several centimeter
 (so-called Schnupp asymmetry) allows the modulation
sidebands to be transferred through the interferometer to the output port,
where they serve as optical local oscillator for the gravitational wave signal.
The photo-current produced by the beat between the different optical
field components (optical demodulation) contains
a radio frequency component at $f_{\rm het} \pm f_{\rm gw}$. In a second demodulation
process the photo-current is then electronically demodulated at $f_{\rm het}$ (using
a mixer) in order to finally derive a signal stream at $f_{\rm gw}$.

In the homodyne readout scheme (center plot of Figure~\ref{fig:principle}) a  small
fraction of the carrier light is split off in front of the interferometer and guided directly
to the output photo detector without passing through the interferometer.
The big
advantage of this form of homodyne readout is that a phase shifter,
 placed in the local oscillator path, allows an easy
change of the optical demodulation phase, i.e.\ the readout quadrature,
without any hardware changes. On the other hand homodyne
readout has the disadvantage that the length and the alignment of the local-oscillator path needs to be highly stable.
In practice this usually implies that the local-oscillator path length as well as its alignment need to be actively
stabilized by a low-noise control system, and all components of the local-oscillator path must be
 seismically isolated inside a vacuum system.
Due to these demanding noise and hardware requirements, so far there have
been no serious plans to change
the readout scheme of the currently operating gravitational wave detectors
from heterodyne to homodyne readout.

DC-readout is a special case of homodyne readout which
is much easier to combine with the existing elements of  currently used
gravitational wave detectors.
In a DC-readout  scheme the operating point of the Michelson interferometer is slightly
shifted off the dark fringe, by introducing a so-called \emph{dark-fringe offset},
thus a certain amount of carrier light leaves the interferometer at the output port
 and can serve as local oscillator. Compared with the previously described homodyne readout,
DC-readout has the advantage that no additional local oscillator path outside the main interferometer
is required. On the other hand, DC-readout offers no easy way to vary the phase of the
optical demodulation.

DC-readout was already used in the first `Michelson' interferometer ever by Michelson and Morley
in 1887 \cite{michelson1887}. It is probably the simplest way to read out a Michelson
interferometer, but was considered to be unsuitable for the first generation of gravitational
wave detectors due to the strong coupling of laser power noise. However, increased
stability of the laser power inside future instruments gives hope for a renaissance
of DC-readout for gravitational wave detectors, which was first proposed
by Fritschel 2000 \cite{Fritschel1, Fritschel2}. A demonstration of a DC-readout
in a suspended prototype interferometer (without signal recycling) has 
recently been performed \cite{rob}.

The next section briefly summarises the general advantages and disadvantages of
DC-readout compared with heterodyne readout, especially taking into account the implications
for an interferometer with tuned or detuned signal recycling \cite{Hild07a}.

\section{Motivation for the use of DC-readout}\label{sec:advantages}

 DC-readout has many advantages over the commonly used heterodyne
readout which are summarized in the following list:
\begin{enumerate}
\item
First of all, when going from the currently-used heterodyne readout scheme to a
\dcr\ scheme the ratio of signal to shot noise will increase \cite{Buonanno03b}. This is
due to the fact that in the homodyne detection the shot noise contribution  from frequencies
 twice the heterodyne frequency
does not exist (please see Section~\ref{sec:shotnoise} for more details).
\item
A reduced number of beating light fields at the detection port potentially reduces and
simplifies the couplings of technical noise \cite{Hild07a}. Especially
the coupling of  amplitude and phase noise of the heterodyne modulation is
strongly reduced in a DC-readout scheme. In addition the frequency noise
coupling to the gravitational wave channel is also expected to be reduced in DC-readout.
%\item
\item
A simpler calibration procedure can be applied, because the GW-signal is present in a single
data-stream even for
 detuned signal-recycling (and not spread over the two heterodyne quadratures as
 described in \cite{Hewitson05}).
\item
As the main photo diode(s) and electronics for the detection do not need to be capable of
 handling RF signals, they can be simplified.
\item
Large-area photo diodes\footnote{RF photo diodes are required to have
a low electrical capacitance.} may be used. These should offer reduced coupling of beam-pointing
 noise, due to decreased beam clipping and decreased influence of 
photo diode inhomogeneity (by averaging over a larger area).
\item
As in the homodyne readout the local oscillator and the GW-signal pass the same optical system an
 optimal spatial overlap is guaranteed. (Due to thermal distortion current GW detectors
 employing arm cavities encountered the problem of imperfect spatial overlap of the carrier light
 (GW signal) and the heterodyne sidebands (local oscillator) \cite{Lawrence03})
\item Finally, the realization of a squeezed light enhanced
interferometer is simpler using DC-readout rather than heterodyne
readout. DC-readout requires squeezed light to be present only at
frequencies in the GW signal bandwidth compared to heterodyne
readout which requires squeezed light around twice the heterodyne
frequency as well \cite{CDRGBM98}. \footnote{ A squeezed light
source working only in the GW signal bandwidth would result in the
DC-readout case in a
sensitivity enhancement limited by the squeezing
strength generated, whereas the same source would act in a
heterodyne-readout based interferometer as if 50\% of the squeezing
was reduced due to losses. Hence, a sensitivity improvement by a
factor of 6\,dB in the DC-readout case would result in the heterodyne
case in an improvement factor of only 2\,dB.}

\end{enumerate}

This long list of advantages has to be compared with the drawbacks of DC-readout.
Even though power fluctuations of the carrier light (i.e.\ the local oscillator) are strongly filtered by the cavity poles
of the power recycling cavity and the high-finesse
arm cavities,  the major disadvantage of DC-readout is (at least for GEO\,600) an
increased coupling of laser power noise (see Section~\ref{sec:sensitivity})\footnote{The
relatively strong coupling of laser power noise in DC-readout was the reason to use
heterodyne techniques instead for the first generation of gravitational wave detectors. However,
with improved stabilisation techniques in recent times, the relative stability of the light inside
the  interferometer is better than the relative phase noise achievable with excellent
 RF techniques.}. In addition there is the potential problem that the response from DC-readout
is not completely linear, due to the operating point sitting on the near-quadratic slope
close to the dark fringe. However, this should not be a significant problem as long as the mean deviation
from the differential arm length operation point is not too large.

\section{Simulated shot noise limited sensitivities of GEO\,600
         for homodyne and heterodyne readout}\label{sec:shotnoise}

\begin{figure}[Htb]
\centering
\includegraphics[width=1\textwidth]{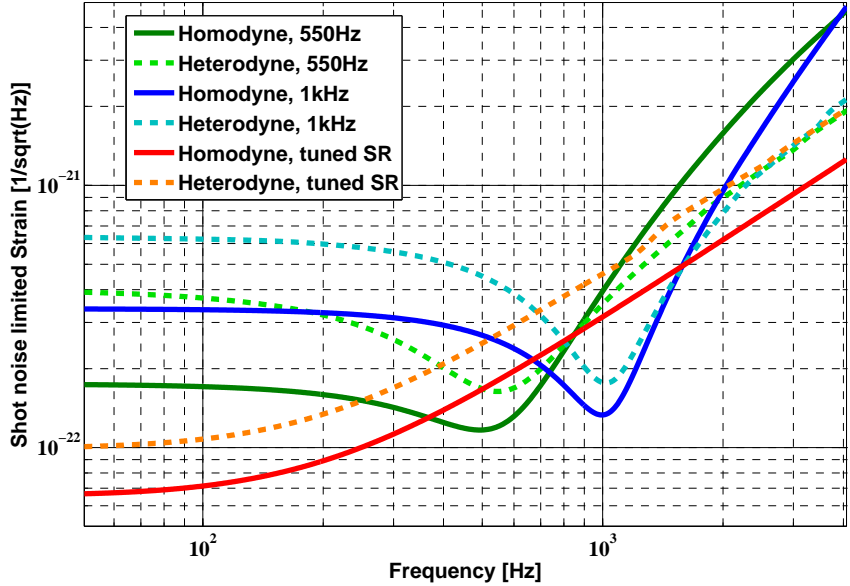}
\caption{Simulated shot noise limited sensitivities of GEO\,600 for three different
SR tuning frequencies, each for heterodyne and \dcr.  } \label{fig:shotnoise}
\end{figure}

Figure \ref{fig:shotnoise} displays the shot noise limited sensitivity
of the GEO\,600 interferometer for heterodyne readout (dashed lines) and DC-readout (solid lines),
 in both cases for each 3 different signal-recycling tunings. These simulations, carried
out with the interferometer simulation software Finesse \cite{finesse}
show two major differences of two different readout methods:
\begin{enumerate}
\item For each signal-recycling tuning the peak sensitivity
achieved with DC-readout is better than the one from heterodyne readout.
\item For detuned signal-recycling the shape of the sensitivity differs for the
two readout methods.
\end{enumerate}

\subsection{Overall shot noise level}
The first point can be explained by a change
of the overall shot noise level. When going from heterodyne readout to
DC-readout the amplitude spectral density of the overall shot noise level decreases
by a factor between $\sqrt{1.5}$ and $\sqrt{2}$
\cite{Niebauer, Meers, Buonanno03b, Harms06}.
%\cite{Buonanno03b}, \cite{Harms06}.
The exact factor  depends on the balancing of the two heterodyne
sidebands at the dark-port. The sensitivity will be increased by
between a factor of $\sqrt{1.5}$ for balanced sidebands and a
factor of $\sqrt{2}$ for completely unbalanced sidebands. This can
 be understood by looking at the individual shot
noise contributions at the dark-port. In the heterodyne case there is additional
shot noise contribution from harmonics of the heterodyne frequency. With 
standard sinusoidal modulation the significant contribution is at twice the modulation
frequency.
%Amongst others we find in
%the heterodyne case shot noise contributions from frequencies
%$2f_{\rm het}$, i.e.\ at twice the heterodyne modulation
%frequency.

\subsection{Shape of the detector response}
\label{sec:dec_shape}
The second point can be explained by a
change of the optical readout quadrature, which determines the shape
of optical gain of the interferometer. The optical gain (or optical transfer function)
 is defined as the transfer
function from differential arm length fluctuations to the error signal of the Michelson
differential arm length servo \cite{sth-status}.

As one can see in Figure~\ref{fig:shotnoise} for detuned signal-recycling,
the shot noise limited sensitivity increases for frequencies below the signal-recycling
tuning frequency and decreases for high frequencies, when going from
 heterodyne readout to DC-readout.
This corresponds to an increase of the optical gain for frequencies below the signal-recycling
tuning frequency and a decrease for high frequencies.
%The change of the optical response, which
%looks similar to a rotation of the overall response shape around the
%point of maximal sensitivity, will be referred to as \emph{rotation of the optical response}.
In  Section~\ref{sec:rotation} we present measurements confirming  this
simulated shape-change of the optical response.

\section{Realisation of DC-readout in GEO\,600}\label{sec:realisation}

\begin{figure}[Htb]
\centering
\includegraphics[width=1\textwidth]{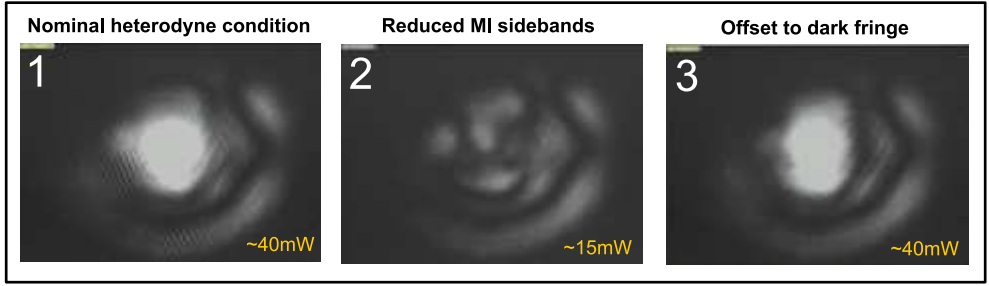}
\caption{Realisation of \dcr\ in GEO\,600. Starting from a fully locked interferometer in
 heterodyne condition (image 1), we first turn down the RF modulation sidebands by a factor
 of 10 (image 2) and finally introduce a dark-fringe offset that couples TEM$_{00}$ carrier
light into the output port, serving as new local oscillator (image 3).} \label{fig:darkport}
\end{figure}
%
%
%Currently the GEO\,600 detector is operated without any dedicated mode cleaner cavity
%at the output port of the interferometer.
%For first tests of DC-readout in the GEO detector it was decided to avoid the
% installation of an output
%mode cleaner, because GEO\,600 was operated in scientific data taking
% mode at that time. The
% installation of a full output mode cleaner system, would have caused not only significant down
%time of the instrument, but more importantly potential risks due to opening the vacuum system
%as well as changing parts of the optical layout of the instrument. The technique used to
%implement DC-readout in GEO\,600 without relaying on an the presence of an output mode
% cleaner  is described in the following.
%
%
%Due to absence of an output mode cleaner the output beam contains not only
%TEM$_{00}$ light, but higher order modes. However, due to the fact that signal-recycling
%provides some \emph{mode-healing}\footnote{In GEO\,600 it was found that the current
%signal-recycling setup, employing a signal-recycling mirror of about 2\,\% transmittance,
%reduces the level of optical higher order modes at the detector output by a factor of 2.\tcr{
%need to check this number...}}, the higher order modes contribution to the output port is only
%about 10 to 15\,mW.
Figure~\ref{fig:darkport}.1 shows an image of the output mode of GEO\,600, when operated
in the normal heterodyne configuration. In the center there is a strong TEM$_{00}$ light field of about
25 to 30\,mW  consisting of the RF modulation sidebands that are transmitted to the
 output via the Schnupp asymmetry and serve as local oscillator for the differential arm
length control system and the gravitational wave readout. In addition one can see some higher mode structure,
which is mostly made of carrier light originating from asymmetric optical imperfections, such as
for instance thermal lensing, inside the main interferometer.

\begin{figure}[Htb]
\centering
\includegraphics[width=0.75\textwidth]{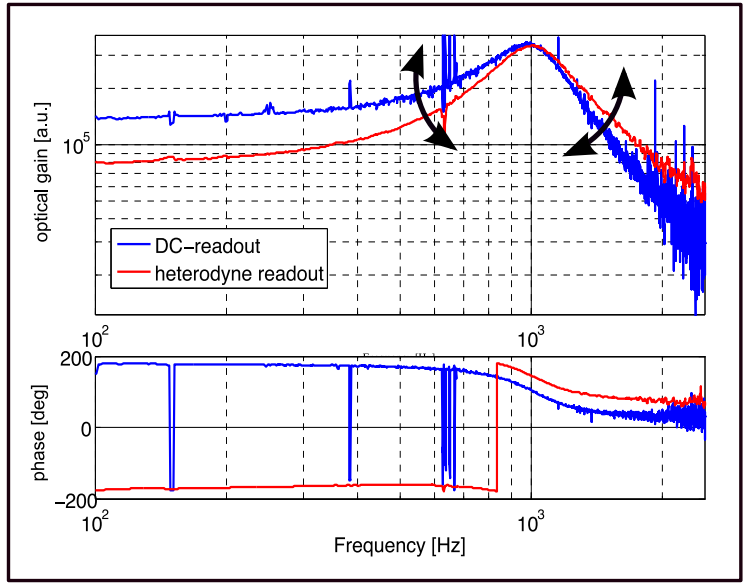}
\caption{Measurement of the optical response of GEO\,600 with detuned signal
recycling (1\,kHz) for DC-readout and heterodyne readout. The change of the optical
transfer function predicted by simulations (see Figure~\ref{fig:shotnoise}) is nicely
confirmed by experimental results.} \label{fig:rotation}
\end{figure}

In order to realize DC-readout of the differential arm length degree of freedom,
 the main task is to replace TEM$_{00}$ RF modulation
 sidebands as local oscillator by TEM$_{00}$ carrier light. We start from  the
interferometer locked in the nominal heterodyne condition and, in a first step, reduce the
amplitude of the RF modulation sidebands by a factor of 10 by turning down the modulation
index in lock, while compensating the gain of all control loops that are affected. In this condition
the optical power at the dark port is reduced to about 15\,mW and  dominated by
higher-order modes (see Figure~\ref{fig:darkport}.2). In a second step we introduce an
offset into the servo electronics controlling the arm length difference in order to dominate
the GEO\,600 output port with TEM$_{00}$ carrier light (see Figure~\ref{fig:darkport}.3).
This procedure of changing from heterodyne to DC-readout has been automated, and 
it takes just a few tens of seconds.\footnote{It has to be noted that only the readout method of the
differential arm length degree of freedom is changed to DC-readout, while most of the auxiliary
length and angular degrees of freedom are still controlled using radio frequency heterodyne 
readout.} 

A big advantage of our method is that no hardware changes are required.
In particular it allows us to establish a DC-readout
scheme without making use of an output mode cleaner which is usually required for
removing the heterodyne sidebands from the light impinging on the main photo diode.
The use of an output mode cleaner would allow the heterodyne sidebands to
be kept at a high level inside the interferometer. 
However, an output mode cleaner is found to be unnecessary in the particular case of GEO\,600, which has good
mode-healing~\cite{Elba-SR}, but is likely to be required on other instruments. 
An
output mode cleaner has the drawback of requiring new hardware together
with a low-noise control system as well as introducing additional  noise sources and
noise couplings.

\section{Comparison of simulated and measured optical response functions}
\label{sec:rotation}

In order to confirm the simulated shape change of the optical response function for GEO\,600
in detuned signal-recycling mode when changing the readout from heterodyne to
homodyne (see Section~\ref{sec:dec_shape})
 we have to measure the optical transfer function for the two different cases.
The result of such a measurement for a signal-recycling tuning frequency of 1\,kHz
is shown in Figure~\ref{fig:rotation}. The shape change of the
optical response predicted by simulations is well confirmed by this measurement.

\begin{figure}[Htb]
\centering
\includegraphics[width=1\textwidth]{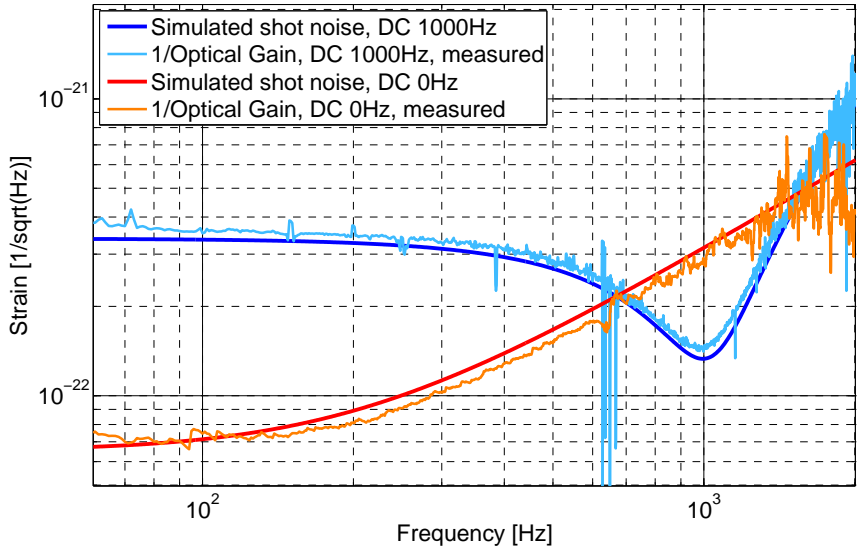}
\caption{Comparison of measured and simulated optical transfer functions of the
GEO\,600 detector with DC-readout for tuned and detuned signal recycling.
The simulation parameters were changed in a predefined way reflecting
the experimental changes. No fitting was applied.
A good agreement of simulated and measured data is found. } \label{fig:sim_vs_meas}
\end{figure}

 In a next set of measurements we compared the optical transfer function of GEO\,600
with DC-readout for tuned and detuned signal recycling. As shown in
Figure~\ref{fig:sim_vs_meas}
the measured optical responses for both detector configurations,
detuned (light blue trace) and tuned signal recycling (orange trace) agree
 accurately with the Finesse \cite{finesse}  simulations (dark blue and red trace). Please note
that this good agreement is achieved using the standard GEO\,600 parameter
set as input for the simulations and no additional fitting or corrections have been applied.

\section{Current sensitivity achieved with DC-readout}
\label{sec:sensitivity}
\begin{figure}[Htb]
\centering
\includegraphics[width=1\textwidth]{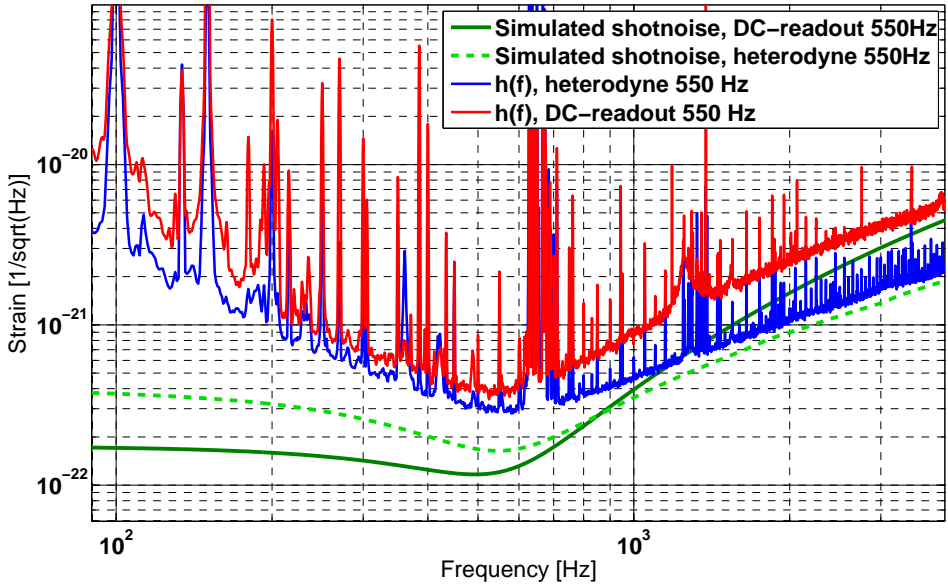}
\caption{Sensitivities achieved with heterodyne readout and DC-readout (solid lines) and
the corresponding simulated shot noise limited sensitivities. All traces represent a signal
recycling detuning of 550\,Hz.} \label{fig:550Hz_all}
\end{figure}

Figure~\ref{fig:550Hz_all} shows a comparison of the nominal GEO\,600 sensitivity
from the fifth LSC science run (blue solid trace) and the sensitivity that was achieved
during first tests of DC-readout (red solid trace) for a signal recycling detuning of 550\,Hz.
 With DC-readout a peak strain sensitivity
of about $4\cdot10^{-22}/\sqrt{\rm Hz}$ at frequencies around 500\,Hz is obtained which
corresponds to a displacement sensitivity of about $2.5\cdot10^{-19}\,{\rm m}/\sqrt{\rm Hz}$.

The solid green and the dashed green traces in Figure~\ref{fig:550Hz_all} represent the
 simulated shot noise limited sensitivity of GEO\,600 for DC-readout and heterodyne
 readout, respectively. As expected from the simulations, at high frequencies the
sensitivity with DC-readout is worse compared to heterodyne readout. However,
the expected benefit from DC-readout, i.e.\ the improved shot noise limited
sensitivity for frequencies below 700\,Hz is found to be covered by excess noises.
Noise projections \cite{Smith-Noise} revealed that the DC-readout sensitivity
is at least partly limited
by laser power noise for frequencies below 300\,Hz \cite{stefan_sydney}. 
In addition it has to be noted that the gap bewteen the simulated shot-noise
limited sensitivity and the measured sensitivity for DC-readout at high frequncies is partly
explained by electronic noise originating from (so far) not optimized detection electronics.
The noise sources limiting the
peak sensitivity with both, heterodyne and DC-readout,
in the frequency band between 300\,Hz and 1\,kHz,
 are so far unexplained and subject of intense investigation.

\section{Summary and Outlook}
\label{sec:summary}

We have developed and implemented a DC-readout scheme at the GEO\,600 gravitational
 wave detector. This scheme  only requires very minor hardware changes, i.e.\ in particular
no output mode cleaner is required for removing the heterodyne sidebands from the
main output port. We compared
the optical response function of GEO\,600 for heterodyne readout and DC-readout
as well as for tuned and detuned signal recycling. The change of the optical response
for detuned signal recycling when going from heterodyne to DC-readout, predicted
by simulations, was well confirmed by experimental data. With DC-readout we
 obtained a strain sensitivity only slightly worse than the nominal GEO\,600 sensitivity
with heterodyne readout. At frequencies around 500\,Hz a peak strain sensitivity of
 $4\cdot10^{-22}/\sqrt{\rm Hz}$ is achieved  which
corresponds to a displacement sensitivity of about $2.5\cdot10^{-19}\,{\rm m}/\sqrt{\rm Hz}$.

These encouraging results have led to the decision to change the nominal readout scheme
of GEO\,600
from heterodyne to DC-readout in Spring 2009. Amongst other hardware changes these
upgrades which will mark the transition from GEO\,600 to GEO-HF \cite{Willke06},
%  ??
will include the installation of a suspended in-vacuum output mode cleaner, whose
main purpose is to remove the light from higher-order optical modes. In the 
medium-future it is planned to operate the GEO detector with tuned signal recycling and DC-readout
in combination with the injection of squeezed light.

\ack{We would like to thank A.~R\"udiger, R.~Schilling and S.~Waldman for many helpful discussions
and comments.
The authors are grateful for support from he Science and Technology Facilities Council
 (STFC) in the UK, the BMBF, Max Planck Society (MPG) and the state of
Lower Saxony in Germany and the European Gravitational Observatory (EGO).
This work was partly supported by DFG grant SFB/Transregio 7 "Gravitational Wave Astronomy".
This document has been assigned LIGO Laboratory document number LIGO-P080105-00-Z.}

\end{document}